\documentstyle[pre,aps,eqsecnum]{revtex}

\input epsf

\begin{document}
\title{A Minimalist Model of Characteristic Earthquakes}
\author{  Miguel~V\'azquez-Prada, Alvaro~Gonz\'alez, Javier~B.~G\'omez and 
Amalio~F.~Pacheco}
\address{Faculty of Sciences, University of Zaragoza,
50009 Zaragoza, Spain.}

\maketitle

\vspace{2cm}

\begin{abstract}
In a spirit akin to the sandpile model of self-organized
criticality, we present a simple statistical model of the
cellular-automaton type which produces an avalanche spectrum
similar to the characteristic-earthquake behavior of some seismic
faults. This model, that has no parameter, is amenable to an
algebraic description as a Markov Chain. This possibility
illuminates some important results, obtained by Monte Carlo
simulations, such as the earthquake size-frequency relation and
the recurrence time of the characteristic earthquake.
\end{abstract}

\section{Introduction}

If there is a well-established fact about regional seismicity this
is the relationship between the magnitude of an earthquake and its
frequency, known as the Gutenberg-Richter (GR) law \cite{gute56}.
This law is of the power-law type when magnitudes are expressed in
terms of rupture area
\begin{equation}
N \propto S^{-b},
\end{equation}
where $N$ is the number of observed earthquakes with rupture area
greater than $S$, and $b$ is the so-called $b$-value, which is a
``universal constant'' in the range 0.8-1.2 \cite{kana75}. The GR
law implies that earthquakes are, on a regional or world-wide
scale, a self-similar phenomenon lacking a characteristic scale (but
see \cite{knop00}).

It is important to notice, however, that the GR law is a property
of {\em regional} seismicity, appearing when we average seismicity
over big enough areas and long enough time intervals. Recently, a
wealth of new data has been collected to extract statistics on
{\em individual} systems of earthquake faults \cite{wesn94}.
Interestingly, it has been found that the distribution of
earthquake magnitudes may vary substantially from one fault to
another and that, in general, this type of size-frequency (SF)
relationship is different from the GR law. Many single faults or
fault zones display power-law distributions only for {\em small}
events, which occur in the intervals between roughly
quasi-periodic earthquakes of much larger size which rupture the
entire fault. These earthquakes are termed ``characteristic'', and
the resulting SF relationship, characteristic earthquake (CE)
distribution.

There is much debate about the origin of the CE distribution
\cite{dahm94}. Because of the short period of instrumental
earthquake records and the scarcity of paleoseismic studies
\cite{wesn94}, the statistics of naturally occurring earthquakes
in single faults are poor. This fact justifies the development of
``synthetic seismicity'' models \cite{robi95}, in which long
catalogs of events are generated by computer models of
seismogenesis. Such models can be tuned by requiring that they
reproduce what is known of the statistics of past seismicity to a
reasonable degree, and then use them to forecast statistical
inferences about the behavior of seismicity using much longer and
homogeneous catalogs of synthetic events.

Many different seismicity models have been presented in the past
twenty years or so. Robinson and Benites \cite{robi95} classify
these modeling approaches into five groups: (1) cellular automata
models, (2) spring-block models, (3) models of single faults in
which slip is discretized into patches and obey simplified
friction laws, (4) continuum models that utilize realistic
constitutive friction laws, and (5) actual physical models.

Cellular automata models \cite{wolf94} have only recently appeared
in  seismological literature, hand in hand with the concept of
self-organized criticality \cite{bak89}. These models are usually
nondeterministic and represent faults as one- or two-dimensional
features. They neglect the details of both elasticity and fault
friction, substituting them by simple cellular automata rules.
Despite their simplicity, they are able to reproduce various types
of SF statistics, including GR and CE distributions \cite{lomn92}.

The key ingredients of any of these models are: (1) the
dimensionality of the fault (1D or 2D), (2) the number of faults
included in the model (one, a few, or many faults), (3) the
employed stress transfer mechanism (nearest-neighbors, long-range
elasticity, mean-field), (4) the degree of incorporation of
inertial effects (quasi-static, quasi-dynamic, or fully dynamic),
(5) the assumed constitutive stress-slip law (experimental,
static-dynamic, velocity-weakening, etc.), and (6) the degree of
stress conservation (conservative versus dissipative models).

Various discrete models of seismicity are able to show a
transition from GR to CE statistics. Among them, we want to cite
here the models of Ceva and Perazzo \cite{ceva93}, Carlson {\em et
al}. \cite{carl93}, Lomnitz-Adler \cite{lomn93}, Rundle and Klein
\cite{rund93}, Ben-Zion and Rice \cite{benz93}, Dahmen {\em et
al.} \cite{dahm94}, Moreno {\em et al.} \cite{more99} and Heinz
and Z\"oller \cite{hein99}. All of them have in common a
transition from a GR to a CE distribution when some of the model
parameters are continuously changed. Some authors interpret the
results in the light of SOC (e.g. \cite{carl93,more99,hein99});
others appeal to percolation theory (\cite{ceva93}); and still
others talk about generic critical point behaviour
(\cite{lomn93,rund93}). But there is no consensus about the origin
of these spatiotemporal complexities, with two extreme views: the
view that says that the spatiotemporal complexity is generated by
the dynamics of the fault, and the opposite view that favors
nonuniformities in geometry and material properties as the cause
of this complexity.

Our purpose here is to build the simplest cellular automaton model
of seismicity capable of displaying a SF relationship of the CE
type. That is, a model which exhibits a power-law relationship for
small events and an excess of big event (of the order of the
system size), together with a very low probability of events of
intermediate size. With respect to the six basic ingredients of
discrete models of seismicity introduced above, the model presented
here is (1) one-dimensional, (2) for a single fault, (3) with a
percolation-like stress-transfer mechanism, (4) quasi-static, (5)
static/dynamic with total stress drop, and (6) dissipative. Also,
because of the inherent simplicity of the model, we want to be
able to derive analytically some of the statistical properties of
the resulting synthetic seismicity using Markov chains.

\section{The Model and its Simulations}

Consider a one dimensional vertical array of  length $N$. The
ordered positions, or levels, in the array will be labeled by an
integer index $i$ varying from $1$ to $N$. This system performs
two functions, it is loaded by receiving  individual particles in
the various positions of the array,  and unloaded by emitting
groups of particles through the first level, $i=1$, which are
called avalanches or earthquakes.

These two functions proceed with the following four rules:
\renewcommand{\labelenumi}{(\roman{enumi})}
\begin{enumerate}
\item  The incoming particles arrive at the system at a constant
time rate. Thus, the time interval between each two succesive
particles will be considered the basic time unit in the evolution
of the system.
\item All the positions in the array, from $i=1$ to $i=N$, have the 
same probability of receiving
a new particle. When a position receives a particle we say that it
is occupied.
\item If a new particle comes to a level which is already occupied, 
this particle
disappears from the system, or in other words, this particle
assignment is wasted. Thus, a given position $i$ can only be
either non-occupied when no particle has come to it, or occupied
when one or more particles have come to it.
\item The level $i=1$ is special. When a particle goes to this first 
position an avalanche
occurs. Then, if all the successive levels from $i=1$ up to $i=k$
are occupied, and the position $k+1$ is unoccupied, the effect of
the avalanche is to unload all the levels from $i=1$ up to $i=k$.
Hence, the size of this avalanche is $k$, and the remaining levels
$i>k$ remain unaltered in their occupancy.
\end{enumerate}

Thus, from what has been mentioned above, this model has no
parameter; the size $N$ is the unique specification to be made,
and the spatial correlation is induced by the iv{\em th} rule.
Now, the state of the system is given by stating which of the
($i>1$) $N-1$ levels are occupied. Each of these states
corresponds to a stable configuration, and therefore the total
number of possible configurations is $2^{(N-1)}$. We use the term
``total occupancy'' for the configuration in which all but the
first level are occupied.

After the occurrence of an avalanche the system is left in a
stable  configuration; the following particle additions go
progressively loading the system, and when a particle is again
assigned to the first level a new avalanche is triggered. Each
avalanche empties the lower levels of the system as explained in
rule (iv), and the system is left in another stable
configuration. The size of the avalanches can range from 1 up to
$N$ and the avalanche of maximum size, $k=N$, will be called the
characteristic one.

From these evolution rules we deduce that after a time unit, i.e.,
after a new incoming particle assignment, we will have an
avalanche if the new particle goes to $i=1$, and this occurs with a
probability $1/N$. Conversely, with a probability of $(N-1)/N$
there will be no avalanche. In this case the system will advance
one unit in its level of occupation when the new particle is
assigned to a non-occupied level, and it will remain at the same
configuration if the assigned level was already occupied.

As this model is 1-dimensional, extensive Monte Carlo simulations
can be performed to accurately explore its properties. In this
paper we focus on two important properties, the avalanche
size-frequency relation and the statistics of the time of return
of the maximum-size avalanche.

\begin{figure}[h]
\epsfxsize=10cm
$$
\epsfbox{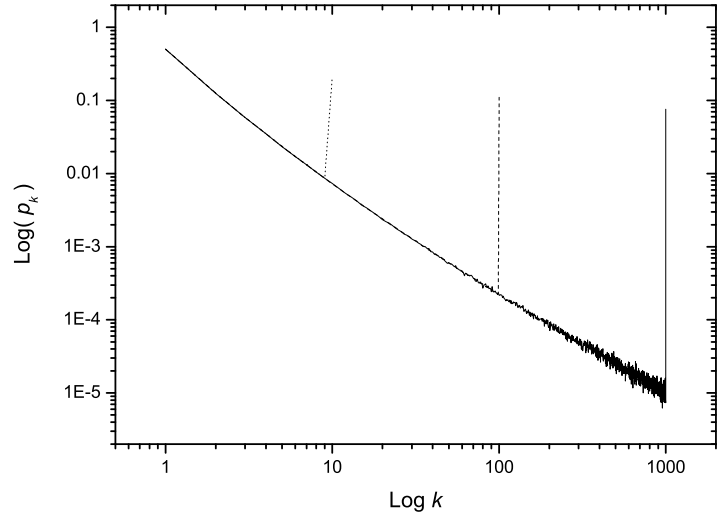}
$$
\caption{Probabiliy of occurrence of earthquakes of magnitude $k$. The 
results for $N=10$,
 $100$, $1000$ are superposed.}
\label{figure1}
\end{figure}

The results for the avalanche size-frequency relation, $p_k$, are
drawn in Fig.~\ref{figure1} and written in Table~\ref{table:1}. In
Fig.~\ref{figure1} we have superposed $p_k$ for $N=10$, $N=100$
and $N=1000$ . This has been partly included in Table~\ref{table:1} as
well. In Fig.~\ref{figure1} there are three notable properties to
be commented on. First and most important, we see that the
characteristic avalanche, $k=N$, has a much higher probability of
occurrence than the avalanches of big size but with $k<N$. In
fact, for $N=10$, $100$ and $1000$, the probability  of their
respective characteristic avalanches does not differ much, and is
near $10\%$. We can express this fact by saying that in this
model, {\em grosso modo}, one would likely only observe very small
avalanches and the characteristic one. Secondly, forgetting for a
moment the case $k=N$, we observe an approximate  power law
behavior, a la Gutenberg-Richter, for the rest of avalanches. The
exponent $b$ of this differential distribution is roughly 1.6. And
thirdly, we observe the perfect coincidence of these curves of
probability for systems of different size $N$. This is also
appreciated in the numbers collected in Table~\ref{table:1}. This,
in a sense, unexpected size-invariance will be discussed in detail
in the next Section.

\begin{figure}[h]
\epsfxsize=10cm
$$
\epsfbox{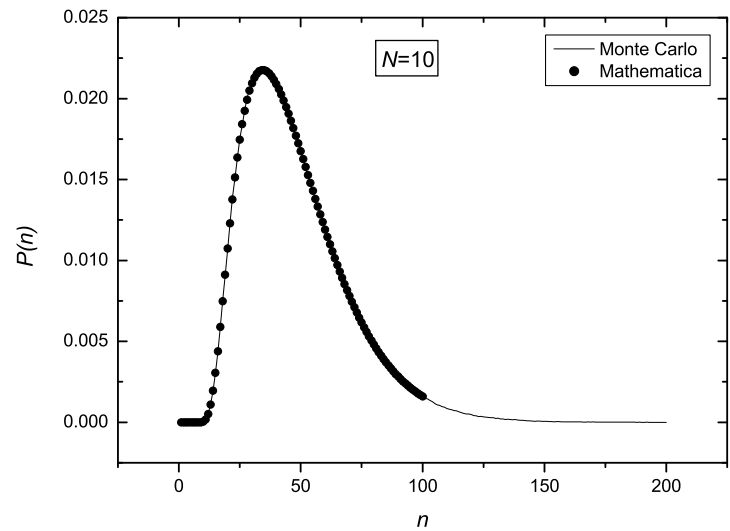}
$$
\caption{Probability of return of the characteristic earthquake as a 
function of time.}
\label{figure2}
\end{figure}

In Fig.~\ref{figure2} we represent the probability curve for ``the
time of recurrence of the characteristic avalanche". This curve,
obtained by Monte Carlo simulations, corresponds to a system of
$N=10$. In the abscissas axis, time (denoted by $n$) starts at 0
just after the occurrence of a $k=N$ avalanche. It is clear in
this model that only after a minimum time lapse of size $N$ the
probability of occurrence of a new $k=N$ avalanche can be non
null. We observe in Fig.~\ref{figure2} that after this minimum
time lapse, $P(n)$ grows to a maximum and then decays. For the
size  $N=10$ analyzed in this figure, the maximum of probability
corresponds to a time interval $n=34$.

\section{The Model as a Markov Chain}

It is easy to become convinced that, for a given $N$, the
$2^{(N-1)}$ stable configurations of our model can be considered
as the states of a finite, irreducible and aperiodic Markov chain
with a unique stationary distribution \cite{durret}. These
configurations are classified in groups according to its {\em
occupation number} (number of occupied levels); the number of
configurations with $j$ occupied levels is $C \left(
 \begin{array}{c}
N-1 \\
j \\
\end{array}
\right)$. One step in the chain corresponds to the result of
adding a new particle to the system. Up to approximately $N= 10$,
the transition matrix, $M$, can be easily obtained using
Mathematica as well as the corresponding stationary probabilities
for each configuration, which correspond to the components of the
eigenvector of $M$ with eigenvalue, $\lambda$, equal to unity.

\begin{figure}[h]
\epsfxsize=10cm
$$
\epsfbox{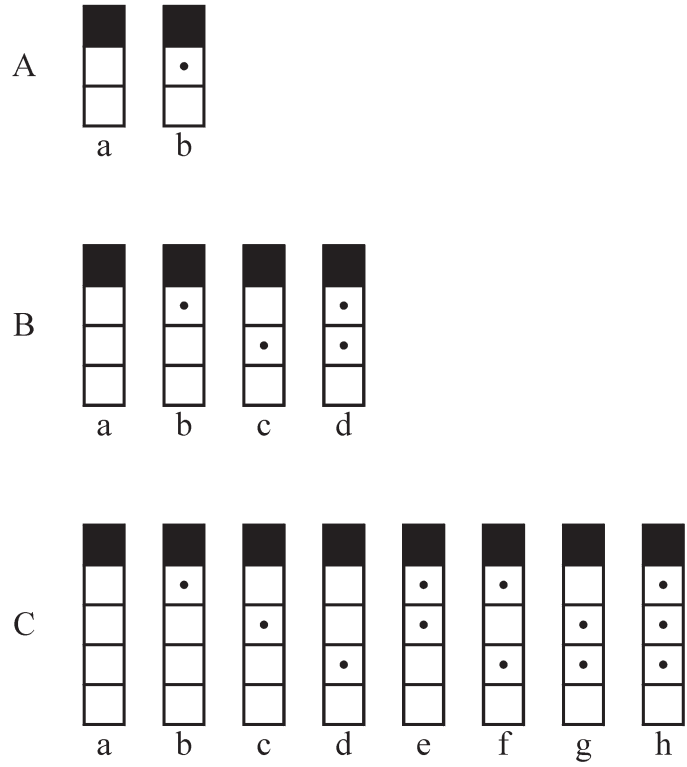}
$$
\caption{Explicit configurations for:  $A) N=2$,   $B) N=3$ and $C)   
N=4$.}
\label{figure3}
\end{figure}

For small $N$, $M$ and its eigenvectors are obtained by
inspection. Let us then start reproducing the first numbers quoted in
Table~\ref{table:1}  for the probabilities of occurrence of
avalanches in systems of small size. With this aim, let us consider
Fig.~\ref{figure3}. There, in (A), (B) and (C) there appear all
the stable configurations, ordered in an increasing state of
occupation, for $N=2$, $N=3$, and $N=4$, respectively. For the
moment, the black level in the top position of the configurations
has no meaning.

For $N=2$, using the same order and notation for the
configurations as in Fig.~\ref{figure3}A, the transition
probabilities  are $M_{a,a} =1/2$, $M_{a,b} =1/2$ , $M_{b,a}
=1/2$, and $M_{b,b} = 1/2$. Thus
\begin{equation}\label{eq:3.1}
M = \left(
\begin{array}{cc}
1/2 & 1/2 \\
1/2 & 1/2 \\
\end{array}
\right),\hspace{5ex} (N=2).
\end{equation}

This $M$ is a doubly stochastic matrix and hence the two stationary 
probabilities
are equal.
\begin{equation}
p_a =1/2, \:\:\: p_b = 1/2,\hspace{5ex} (N=2).
\end{equation}
For $N=3$, the non-null transition probabilities are: $M_{a,a} =
1/3$, $M_{a,b} =1/3$, $M_{a,c} =1/3$; $M_{b,b} =2/3$, $M_{b,d}
=1/3$; $M_{c,a}=1/3$, $M_{c,c} =1/3$, $M_{c,d} =1/3$; $M_{d,a}
=1/3$, and $M_{d,d} =2/3$. Thus
\begin{equation}\label{eq:3.2}
M = \left(
 \begin{array}{cccc}
1/3 & 1/3 & 1/3 & 0\\ 0 & 2/3 & 0   & 1/3\\ 1/3   & 0   & 1/3 &
1/3\\ 1/3 & 0   & 0   & 2/3\\
\end{array}
\right),\hspace{5ex} (N=3).
\end{equation}
And the components of the eigenvector corresponding to $\lambda=1$ are:
\begin{equation}
p_a=1/4, \:\:\: p_b=1/8,  \:\:\: p_c=1/4,  \:\:\: p_d=
3/8,\hspace{5ex} (N=3).
\end{equation}

Finally, for $N=4$ the non-null transition probabilities are
$M_{a,a}=1/4, M_{a,b}=1/4, M_{a,c} =1/4,  M_{a,d} =1/4;
M_{b,b} =2/4, M_{b,e} =1/4, M_{b,f} =1/4; M_{c,c} = 2/4, M_{c,e} =1/4, 
M_{c,g} =1/4,
M_{d,a} =1/4,  M_{d,d} =1/4, M_{d,f} =1/4, M_{d,g} =1/4; M_{e,e} =3/4, 
M_{e,h} =1/4;
M_{f,b} =1/4, M_{f,f} =2/4, M_{f,h} =1/4; M_{g,a} =1/4, M_{g,g} =2/4, 
M_{g,h} =1/4;
M_{h,a} =1/4, M_{h,h} =3/4$. Thus

\begin{equation}
M = \left(
 \begin{array}{cccccccc}
1/4 & 1/4 & 1/4 & 1/4 & 0 & 0 & 0 & 0 \\
0 & 2/4 & 0 & 0 & 1/4 & 1/4 & 0 & 0 \\
0 & 0   & 2/4 & 0 & 1/4 & 0 & 1/4 & 0 \\
1/4 & 0 & 0 & 1/4 & 0 & 1/4 & 1/4 & 0 \\
0 & 0 & 0 & 0 & 3/4 & 0 & 0 & 1/4 \\
0 & 1/4 & 0 & 0 & 0 & 2/4 & 0 & 1/4 \\
1/4 & 0 & 0 & 0 & 0 & 0 & 2/4 & 1/4 \\
1/4 & 0 & 0 & 0 & 0 & 0 & 0 & 3/4 \\
\end{array}
\right),\hspace{5ex} (N=4).
\end{equation}
And, after its diagonalization, one finds
\begin{eqnarray}
& p_a = 9/64,  \:\:\: p_b = 7/64,  \:\:\: p_c =9/128,  \:\:\: p_d
=3/64, \nonumber \\ & p_e =23/128,  \:\:\: p_f = 5/64,  \:\:\: p_g
= 15/256, \:\:\: p_h = 81/256,\hspace{5ex} (N=4).
\end{eqnarray}
From these numbers, one deduces that in a system with $N=2$ levels
one should expect avalanches of size $k=1$ with a probability $p_1
= p_a =1/2$ , and of size $k =2$ with $p_2 = p_b =1/2$. In $N=3$,
$p_1 = p_a + p_b = 1/2$,  $p_2 = p_c =1/8$, and $p_3 = p_d =3/8$.
And in $N =4$, $p_1 = p_a+ p_b + p_c +p_e =1/2$,  $p_2 = p_d + p_f
= 1/8$, $p_3 = p_g = 15/256 = 0.05859..$, and $p_4 = p_h =
81/256$.

Thus, we have observed that in systems with $N=2$, $N=3$ and $N=4$
levels, the value of $p_1$ is a constant equal to $1/2$ and this
result coincides with Table~\ref{table:1}. We have also observed
that in $N=3$ and $N=4$ the value of $p_2$ is a constant equal to
$1/8$, and in $N =4$ we have deduced that $p_3= 0.05859$. All
this agrees with Table~\ref{table:1}.

A conclusive argument  proving that in this model the value of
$p_k$ is a constant independent on the size $N$ is obtained by
re-analyzing Fig.~\ref{figure3} from a wider perspective. Now we
consider that in Fig.~\ref{figure3}A, the configuration labeled by
$a$ represents, in a system of size $N$,  all the configurations
in which the level $i = 2$ is not-occupied; the rest of the levels
$i>2$ are in any possible state of occupation (this is represented
by the black level at the top of the configuration). And in 
Fig.~\ref{figure3}A, configuration $b$, we consider a system of
size $N$ which has its second level occupied. The probabilities
of these two cases must add to unity. Then defining a Markov chain
for these two excluding states, and using the same notation as
before, we find $M_{a,a} = (N-1)/N, M_{a,b} = 1/N, M_{b,a} =1/N$,
and $M_{b,b} =(N-1)/N$. The diagonalization of this matrix leads
for $p_a$ and $p_b$ to the same results written in
Eq.~(\ref{eq:3.2}), $p_a=p_b=1/2$; however, its interpretation now
is different. Here, $p_a=1/2$ implies that for any value of $N$,
the probability of having an avalanche $k=1$ is 1/2;  and
$p_b=1/2$ simply means that the probability of having avalanches
$k>1$ is 1/2. Using the same line of reasoning, and referring to
Fig.~\ref{figure3}B, configuration $a$ represents all the
configurations, in a system of size $N$, where levels $i=2$ and
$i=3$ are not occupied. And configuration $b$ represents all the
configurations where the level $i =2$ is free and the level $i =3$
is occupied, etc. Then, the non null transition probabilities are
$M_{a,a} = (N-2)/N, M_{a,b} = 1/N, M_{a,c} =1/N; M_{b,b} =
(N-1)/N, M_{b,d} = 1/N; M_{c,a} =1/N, M_{c,c} =(N-2)/N, M_{c,d}
=1/N; M_{d,a} =1/N, M_{d,d} =(N-1)/N$. The diagonalization of this
matrix provides the same stationary probabilities quoted in
Eq.~(\ref{eq:3.2}). Here $p_c=1/8$ means that for an arbitrary
$N$, the probability of occurrence of avalanches of size 2, $p_2$
is $1/8$. And the fact that $p_a+p_b =1/2$ confirms that for any
$N$,  $p_1=1/2$.

Extending this line of reasoning to the 8 configurations drawn in
Fig.~\ref{figure3}C, one concludes that, for any $N$, $p_3 = pg
=15/256=.058593..$ and one verifies the previous conclusions
$p_1=1/2$ and $p_2=1/8$.

Therefore, in this model, if for the system of size N one knows all the
$p_k$ from $k=1$ to $k=N$, then for the system of size $N+1$  the $p_k$ 
are identical ,
with the exception of the last two, and these  fulfill
\begin{equation}
p_N(N) = p_{N+1}  (N+1) +p_N  (N+1)
\end{equation}
the recursive way in which $p_N(N)$ divides into $p_{ N+1} (N+1)$
and $p_N  (N+1)$ is, however, non trivial.

Let us analyze now, from the Markov-chains point of view, the
results for the time of return of the characteristic earthquake
shown in Fig.~\ref{figure2}. After a   $k =N$ avalanche, the
system is left in the configuration of no occupancy (for the
present discussion we will refer to this configuration as $a_1$).
A new characteristic avalanche will occur when, starting from the
configuration $a_1$, the system reaches the configuration of total
occupancy (which will henceforth be denoted by $a_N$), and then
the next particle is assigned to the $i=1$ level. The number of
time steps elapsed between $a_1$ and $a_N$, plus 1, will be
denoted by $n$. And our purpose is to compute the probability of
occurrence of a $k=N$ avalanche as a function of $n$.  It must be
understood that between $a_1$ and the occurrence of the next $k=N$
avalanche  on time $n$, the system may have visited $a_N$ an
arbitrary number of times but without triggering any  $k=N$
avalanche. In other words, in those visits there has been no
transition from $a_N$ to $a_1$.

In Markov chains, the transition matrix $M$ gives the probability
of going from one configuration to another in one step, and the $m$
step transition probability is the $m$-th power of $M$. Thus a
simple way to compute $P(n)$ is  the following:
\renewcommand{\labelenumi}{\arabic{enumi}}
\begin{enumerate}
\item  Take  $M$, point to the element in the last row and the first 
column,
and substitute it by 0. We will denote the new matrix $M'$.

\item Compute $T_n = {M'}^{(n-1)}$.
\item  Take the element of the first row and the last column of $T_n$, 
and multiply it
by $1/N$. This  result is  $P(n)$.
\end{enumerate}

This is clear, because $M'$ does not permit transitions from $a_N$
to $a_1$. Thus, in $T_n$ we have all the probabilities of
transitions, in $n-1$ steps, between all the configurations, with
the restriction  that from $a_N$ to $a_1$ this transit is
forbidden. Hence, in $T_n$ the matrix element of the first row and
the last column corresponds to the transition from $a_1$ to $a_N$,
in $n-1$ steps and with the mode $a_N \longrightarrow a_1$ locked.
Finally, as the $1/N$ factor is the probability of the transition
$a_N \longrightarrow a_1$,
 we actually have built $P(n)$.
In Fig.~\ref{figure2}, for $N=10$, we see the perfect match
between the Monte Carlo simulations and the results coming the
theory of Markov, calculated using Mathematica.

In order to get a quantitative insight on $P(n)$, let us apply
this method to $N = 2$. In this case, $M$ is given by
Eq.~(\ref{eq:3.1}). Hence,
\begin{equation}
M'= \left(
 \begin{array}{cc}
1/2 & 1/2 \\
0   & 1/2 \\
\end{array}
\right),  \hspace{3ex}
 M'^{(n-1)} =  \left(\frac{1}{2}\right)^{(n-1)}
\left(
 \begin{array}{cc}
1 & n-1 \\
0 & 1 \\
\end{array}
\right),
\end{equation}
and, therefore,
\begin{equation}
P(n) = (n-1)\cdot (1/2)^{(n-1)}\cdot (1/2) = \frac{n-1}{ 2^n},
\hspace{5ex} (N=2).
\end{equation}
Thus, we observe that the asymptotic fall-off behavior of $P(n)$
is of the type
\begin{equation}
P(t) \propto t \exp(-t), \hspace{5ex} (N=2).
\end{equation}
It is important to recall that in aperiodic, irreducible and
finite Markov Chains such as that of our model,  the  mean waiting time
for a configuration is the inverse of the stationary probability
of that configuration. Then, the mean time between characteristic
avalanches in this model is
\begin{equation}
      <n> = \frac{1}{\frac{a_N}{N}} = \frac{N}{a_N}.
\end{equation}

As an example, for $N=4$,  $<n> = 4 \cdot (256/81) = 12.4$ time units.

\section{CONCLUSIONS}

We have presented a one-dimensional discrete model of seismicity
that displays a size-frequency spectrum similar to that expected
from a characteristic-earthquake behavior. Although this
stochastic model  obviously lacks the explicit physics of other
detailed --and complex--  dynamical models of earthquakes, its
basic hypotheses and implications are clear, phenomenologically
reasonable and coherent. The size invariance of this model, also
surprisingly manifests itself in predicting an identical
probability of occurrence for earthquakes of the same magnitude
independently of the size $N$ of the system. In this universal
rule the characteristic earthquake is excluded. This model has the
additional bonus that several important predictions can be
algebraically derived by using the theoretical framework of the
Markov chains. Specifically, the statistics of the time of return
of the characteristic earthquake are neatly predicted by this
formalism.

Dahmen et al.~\cite{dahm94} report a model able to transit from
the GR to the CE behavior and, what is more interesting, these
authors define a ``configurational entropy" as an appropriate
concept that reflects in which of these two extreme behaviors  the
system is actually operating.  In qualitative terms, the GR
behavior corresponds to a high entropy mode of operation while the
CE behavior corresponds to a low entropy mode. Therefore, we would
like to make an assesment of our model from this configurational
entropy point of view and check the concordance, or not, with the
conclusions of Ref.~\cite{dahm94}.

\begin{figure}[h]
\epsfxsize=10cm
$$
\epsfbox{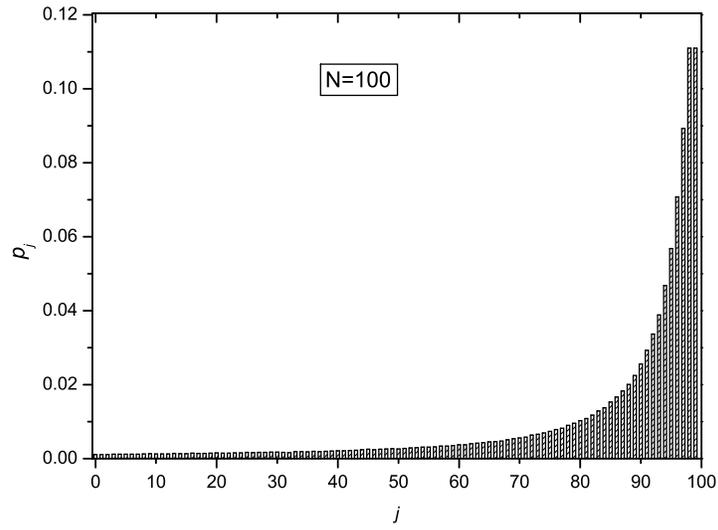}
$$
\caption{Stationary probabilities as a function of the number of levels 
occupied, $j$.}
\label{figure4}
\end{figure}

In our model, the configurations are classified in groups
according to  the number of levels, $j$,  that are occupied.  The
statistical weight of each $j$ is $C \left(
 \begin{array}{c}
N-1 \\
j \\
\end{array}
\right)$, which has its maximum values for $j$ around $N/2$.
Conversely, the statistical weight is minimum on  the extrema: for
$j$ around $1$ and for $j$ near $N-1$. Then, we need to find out
the values of the stationary probabilities of each occupation
number $j$. This is shown in Fig.~\ref{figure4} for a system of
size $N=100$. There we observe how in our model the system resides
most of the time in the configurations of maximum occupancy, that
is, where the  configurational entropy is a minimum, agreeing with
the interpretation of Ref.~\cite{dahm94}.

As a final minor remark, note that in Fig.~\ref{figure4}
$p_{(j=N)}$ and $p_{(j=N-1)}$ are identical. This is a property
that holds in this model for any value of $N$ and which can be
easily proved. For brevity reasons, we omit this proof.



\acknowledgments{ A.F.P. thanks Jes\'us As\'{\i}n, Jorge
Ojeda and Carmen Sang\"uesa for many fruitful discussions. This
work was supported by the Spanish DGICYT (Project PB98-1594}).


\newpage
\begin{table}[h]
\begin{tabular}{|d|ddddd|}
$k$ &$N=2$ & $N=3$ & $N=4$     & $N=10$         & $N=100$   \\ \hline
1 & 0.5  &  0.5    & 0.5       &    0.500022    & 0.49992 \\
2 & 0.5  & 0.125   & 0.125     &   0.124974     & 0.12513\\
3 &     &  0.375   & 0.0585938 &   0.058599     & 0.0586\\
4 &     &          & 0.316406  & 0.034855       & 0.03486\\
5 &     &          &           &  0.023499      & 0.02346\\
6 &     &          &           &    0.017126    & 0.01714\\
7 &     &          &           &    0.013151    & 0.01313\\
8 &     &          &           &    0.010506    & 0.01051\\
9 &     &          &           &    0.008627    & 0.00864\\
10 &    &          &           &    0.208636    & 0.00721\\
99 &    &          &           &                & 2.28507 $\cdot 
10^{-4}$\\
100&    &          &           &                & 0.11132\\
\end{tabular}
\caption{Probability of occurrence of the earthquake of magnitude
$k$ for a different system size, $N$.}
\label{table:1}
\end{table}

\end{document}